\documentclass[twocolumn, aps, prl]{revtex4-2}

\usepackage{xcolor,graphicx}%,tikz}
\usepackage[colorlinks=true]{hyperref}
\usepackage{braket}
\usepackage[retainorgcmds]{IEEEtrantools}
\usepackage{lineno}

\DeclareUnicodeCharacter{2009}{\,} 

\begin{document}
\title{Deterministic Storage and Retrieval of Telecom Quantum Dot Photons Interfaced with an Atomic Quantum Memory}
\author{S. E. Thomas$^{1}$\hyperlink{equalcontributions}{$^\dagger$}, L. Wagner$^{2}$\hyperlink{equalcontributions}{$^\dagger$}, R. Joos$^2$, R. Sittig$^2$, C. Nawrath$^2$, P. Burdekin$^1$, T. Huber-Loyola$^3$, S. Sagona-Stophel$^1$, S. H\"{o}fling$^3$, M. Jetter$^2$, P. Michler$^2$, I. A. Walmsley$^1$, S. L. Portalupi$^2$, P. M. Ledingham$^4$}
\email{P.Ledingham@soton.ac.uk}
\affiliation{$^1$QOLS, Department of Physics, Imperial College London, London SW7 2BW, UK\\
$^2$Institut für Halbleiteroptik und Funktionelle Grenzflächen (IHFG), Center for Integrated Quantum Science and Technology (IQ$^{\text{ST}}$) and SCoPE, University of Stuttgart, Allmandring 3, 70569 Stuttgart, Germany\\
$^3$Julius-Maximilians-Universität Würzburg, Physikalisches Institut and Würzburg-Dresden Cluster of Excellence ct.qmat, Lehrstuhl für Technische Physik, Am Hubland, 97074 Würzburg, Germany\\
$^4$Department of Physics and Astronomy, University of Southampton, Southampton SO17 1BJ, UK}
\date{\today}
	
\begin{abstract}
A hybrid interface of solid state single-photon sources and atomic quantum memories is a long sought-after goal in photonic quantum technologies. Here we demonstrate deterministic storage and retrieval of photons from a semiconductor quantum dot in an atomic ensemble quantum memory at telecommunications wavelengths. We store single photons from an InAs quantum dot in a high-bandwidth rubidium vapour based quantum memory, with a total internal memory efficiency of $(12.9 \pm 0.4) \%$. The  signal-to-noise ratio of the retrieved photons is $18.2\pm 0.6$, limited only by detector dark counts. This demonstration paves the way to quantum technologies that rely on distributed entanglement, and is especially suited for photonic quantum networks.
\end{abstract}
\maketitle

A key requirement for the future quantum internet is the distribution of entanglement between remote nodes of a quantum network~\cite{Kimble2008, Wehnher2018}. Semiconductor quantum dots (QDs) have emerged as high-performing on-demand sources of single~\cite{Somaschi2016a} and entangled photons~\cite{Huber2017}. In order to distribute entanglement across a quantum network, it is necessary to have efficient high-fidelity storage and on-demand recall of these single photon states in a quantum memory~\cite{Heshami2016}, and operating within the low-loss telecommunication bands of optical fibres is critical for scaling quantum repeaters to global distances. Such quantum repeater infrastructures based on QDs and atomic-ensemble-based quantum memories~\cite{Neuwirth_2021} can result in entanglement distribution rates that outperform DLCZ-type protocols~\cite{Duan2001,Simon2007d} by orders of magnitude~\cite{Minar2012, Sangouard2012}. The storage and on-demand recall of single photons from a semiconductor quantum dot into a quantum memory is therefore a key goal in photonic quantum technologies.

In order to achieve an efficient interaction between a quantum dot photon source and a quantum memory, both the central wavelength and the bandwidth of the emitted photon are required to be matched to the memory operation frequency and bandwidth. Considerable progress has been made towards fabricating QDs that emit light with wavelengths close to atomic transitions. Fine-tuning of the emitted wavelength can then be achieved through the application of external electric~\cite{Bennett2010} or magnetic fields~\cite{Akopian2011}, strain via piezoelectric actuators~\cite{Kuklewicz2012,Jons2012,Trotta2012}  or by temperature~\cite{Kiraz2001}. The significant challenge arises from matching the quantum memory bandwidth to that of the GHz-bandwidth QD emission. Many of the leading memories are incompatible, operating in the near MHz regime~\cite{Hedges2010, Cho2016, Vernaz-Gris2018}. A promising approach is to use two-photon absorption mechanisms in hot alkali vapours. Electromagnetically-induced transparency (EIT)~\cite{Wolters2017, Buser2022} and the fast ladder memory (FLAME)~\cite{Finkelstein2018} have pushed the operational bandwidth to the near-GHz regime. The Raman~\cite{Michelberger2015} and off-resonant cascaded absorption (ORCA) memories have achieved bandwidths of $1\,$GHz~\cite{Kaczmarek2018, Gao2019}. 

While challenging, there has been considerable progress towards interfacing quantum dot photons with atomic systems. Light from QDs has been shown to interact with hot rubidium~\cite{Akopian2011, Huang2017} and caesium vapour~\cite{Wildmann2015, Trotta2016, Portalupi2016, Vural2018, Kroh2019, Maisch2020} showing slow- and fast-light effects, or dispersion-induced delay,  at near-infrared wavelengths. Further, the temporal intensity profile of QD photons has been observed to be modified by fast EIT in a hot rubidium cell~\cite{Cui2023}. Direct coupling of $935\,$nm QD light with a trapped Yb$^+$ ion facilitated by a high-finesse fibre-based optical cavity has been demonstrated~\cite{Meyer2015}. Finally, dispersive delay of $879.7\,$ nm QD light in a cryogenically-cooled Nd$^{3+}$:YVO$_{4}$ crystal has been achieved using the atomic frequency comb protocol \cite{Tang2015}. However, storage and on-demand retrieval of photons from a quantum dot single-photon source in an atomic quantum memory has not previously been demonstrated, owing to the key challenges of quantum memory efficiency, bandwidth and noise. 

\begin{figure*}[t]
\centering\includegraphics[width=\linewidth]{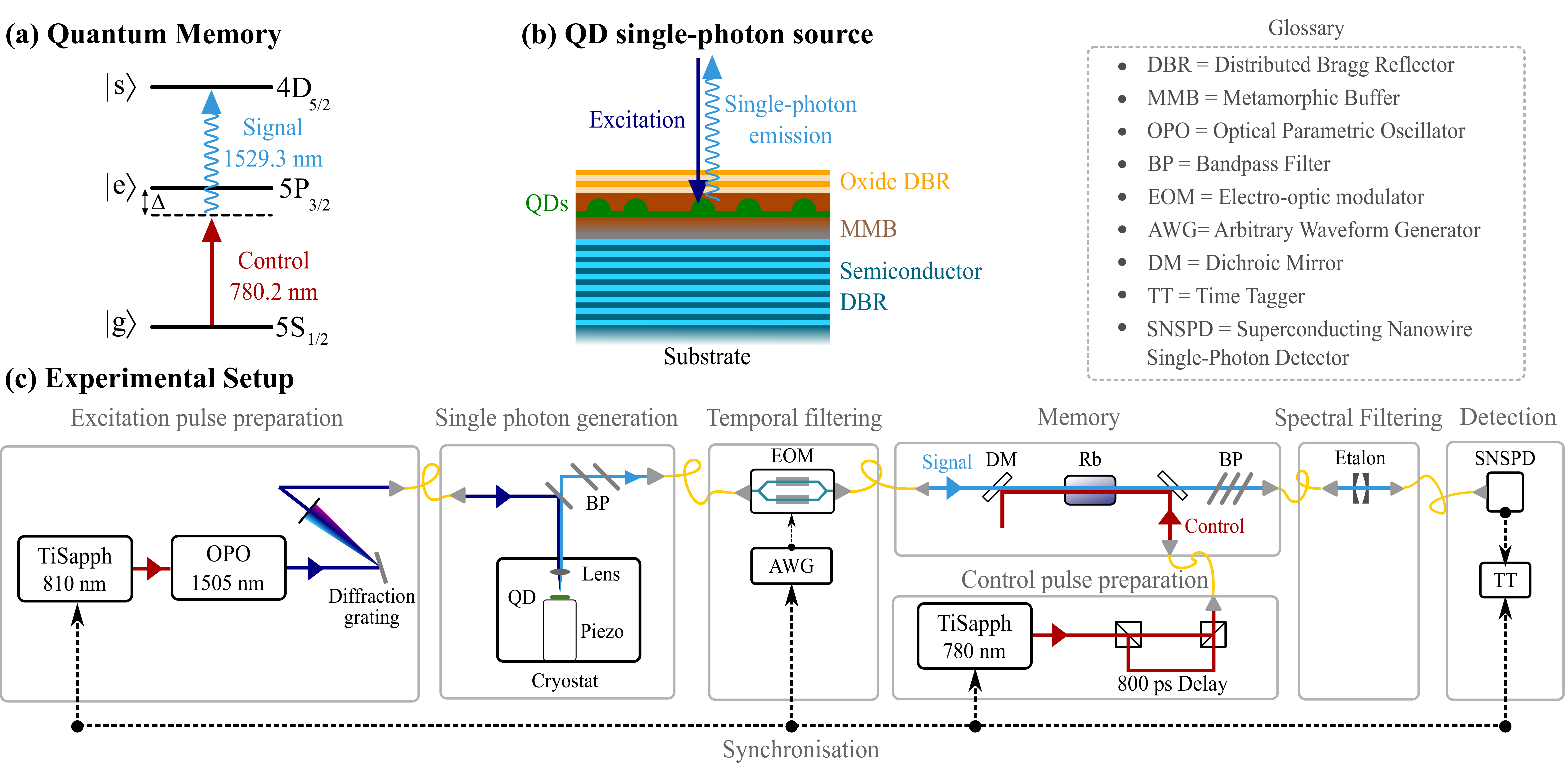}
\caption{\textbf{Schematic of the experimental set-up for the quantum dot - quantum memory interface.} (a) Energy level scheme for the Telecom ORCA quantum memory protocol in rubidium vapour. (b) Scheme of the semiconductor QD sample with semiconductor bottom DBR, metamorphic buffer (MMB) and oxide top DBR. 
(c) Experimental set-up of the hybrid interface to store photons from a quantum dot single-photon source in a quantum memory.  \label{fig:setup}}
\end{figure*}

In this work, we experimentally demonstrate the storage and on-demand recall of deterministic telecom single photons emitted from a semiconductor QD in a hot rubidium vapour quantum memory. This is achieved using the ORCA memory protocol which was recently demonstrated to operate at telecommunication wavelengths with an excellent unconditional noise floor of less than $10^{-6}$ noise photons per pulse and a $1\,$GHz bandwidth~\cite{Thomas2022}. The single photons are delivered by a MOVPE-grown InAs QD emitting at $1529.3\,$nm with GHz bandwidth~\cite{Sittig2022}, which is optimised to wavelengths close to that of the ORCA memory. Spectral and temporal shaping are used to further condition the light to match the properties of the memory. To our knowledge this is the first demonstration of storage and deterministic retrieval of photons from a quantum dot single-photon source in an atomic quantum memory -- a hybrid quantum light-matter interface. Furthermore, our demonstration is at telecommunication wavelengths which is critical for scalability of quantum networks. This demonstration marks a significant step toward practical implementations of photonic quantum technologies.

\section{Results}

The quantum memory is based on the ORCA protocol in rubidium vapour, where a strong control pulse detuned from the $780.2\,$nm Rb D2 line $\left(5\mathrm{S}_{1/2} - 5\mathrm{P}_{3/2}\right)$ by $\Delta$ dynamically induces the absorption of the QD photon at $1529.3\,$nm into the $4$D$_{5/2}$ excited state, as shown in Figure~\ref{fig:setup}(a). The QD photon is then stored as a collective coherence between the ground and storage states. The ORCA memory protocol offers on-demand retrieval, since the application of a second control pulse at a controllable later time maps this atomic coherence back to an optical field. 

\begin{figure}
\centering\includegraphics[width=\linewidth]{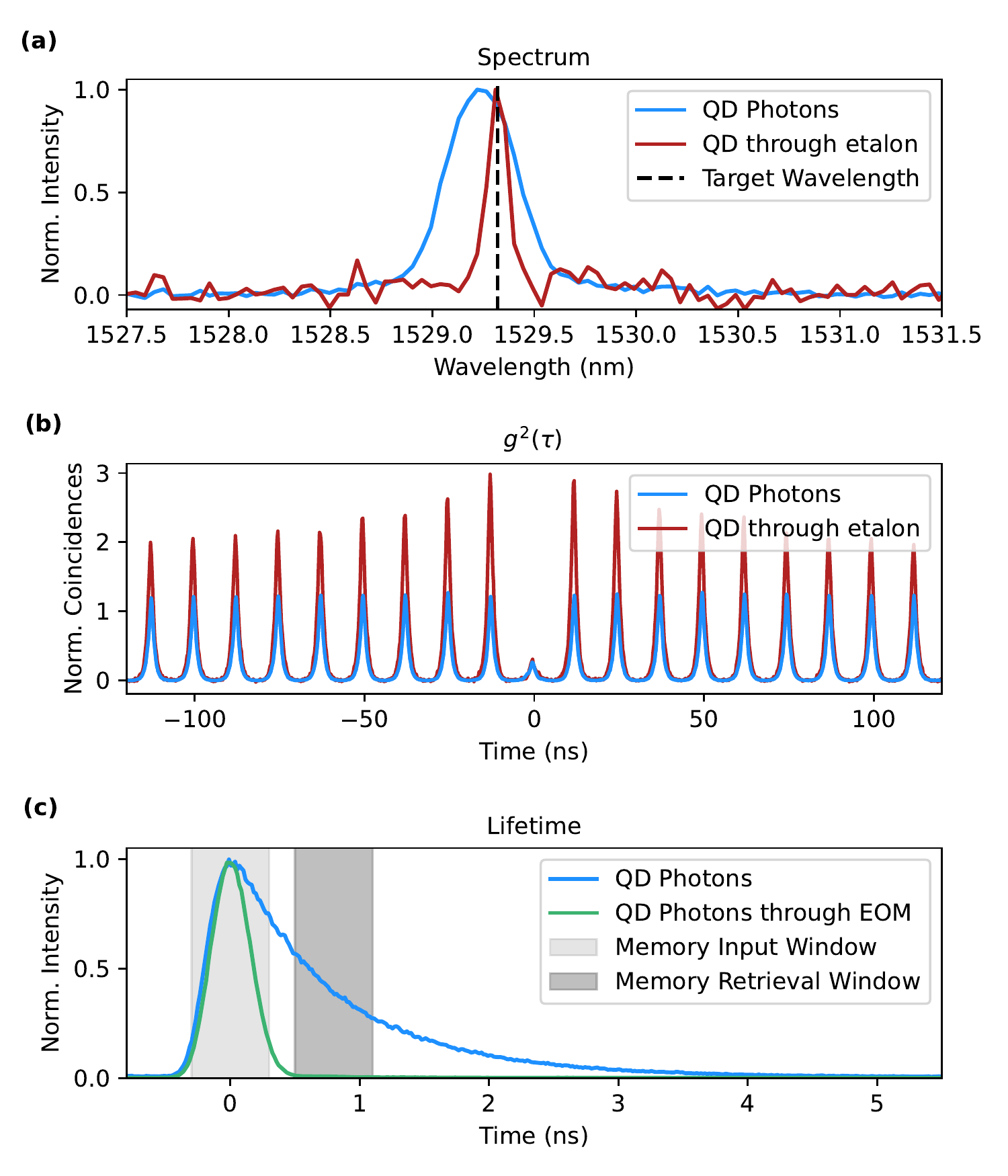}
\caption{ \textbf{Characterisation of single-photon source} (a) Spectrum of photons from QD without (blue) and with (red) spectral filtering from the etalon, normalised to the maximum intensity. The dashed black line shows the target wavelength of the memory. (b) The second-order autocorrelation, $g^{(2)}(\tau)$, of photons from the QD without (blue) and with (red) the etalon. The data is normalised to the height of the Poissonian coincidence peaks at long time delays. The etalon introduces blinking which results in bunching (increased coincidences) at short time delays (see Supplementary Material). (c) Arrival time histogram of photons from the QD without (blue) and with (green) temporal filtering through the EOM, normalised to the maximum intensity.  The shaded regions indicate the input and retrieval windows of the quantum memory, with a chosen storage time of 800~ps.  \label{fig:QDcharacterisation}} 
\end{figure}

 The single-photon source is based on InAs quantum dots (QD) grown on top of a jump-convex-inverse InGaAs metamorphic buffer (see Figure~\ref{fig:setup}(b)) which enables emission in the telecom-band~\cite{Sittig2022}. The QDs are grown {inside a cavity formed by a bottom distributed Bragg reflector (DBR) comprised of $23$ AlAs/GaAs pairs and a top DBR of $4$ SiO$_2$/TiO$_2$ pairs (see Methods). This planar cavity structure increases the QD light extraction, and the thickness is adapted for operation at $1529.3\,$nm.} The QD is excited through quasi-resonant p-shell excitation using 3~ps laser pulses with a central wavelength of 1505~nm and a repetition rate of 80~MHz. The single photons are separated from the excitation laser using three bandpass filters with a 12~nm FWHM bandwidth. A high-resolution spectrometer is used to identify a QD which emits light at precise wavelength of the ORCA quantum memory, as shown in Figure~\ref{fig:QDcharacterisation}(a). The output of the QD single-photon source is measured using superconducting nanowire single photon detectors (SNSPDs), and the recorded count rate is $4 \times 10^5 \, \mathrm{s}^{-1}$. The second-order autocorrelation of the collected light is measured in a Hanbury Brown-Twiss set-up, and found to be $g^{(2)}(0) = 0.306 \pm 0.002 $ (see Figure~\ref{fig:QDcharacterisation}(b)), { which verifies non-classicality and is consistent with a predominantly single photon output.  }

In order to efficiently interface the QD single-photon source with the quantum memory we perform spectral and temporal filtering of the single photons. The telecom ORCA memory has a characteristic lifetime of 1.1~ns~\cite{Thomas2022} over which the memory efficiency decays, so in order to efficiently retrieve photons from the memory we choose a retrieval time of 800~ps. The single photons have an exponentially-decaying temporal intensity profile with a $1/e$ decay time of $0.85\pm 0.01$~ns, and therefore there is a significant probability of input photons in the retrieval window of the memory, as shown in Figure~\ref{fig:QDcharacterisation}(c). To ensure that the input and retrieval signals are temporally well-separated, the single photons are temporally filtered to a Gaussian temporal mode shape with a full-width half maximum (FWHM) of 300~ps, as shown in Figure~\ref{fig:QDcharacterisation}(c). This is achieved using an EOM intensity modulator that is driven by an arbitrary waveform generator. 

The operational  bandwidth of the quantum memory is determined by the bandwidth of the control pulse which is 1~GHz. The homogeneous linewidth of a single photon emitted from the QD is around $200\,$MHz as estimated from the lifetime in Figure~\ref{fig:QDcharacterisation}(c). However, due to charge noise and spin noise in the system, the central frequency of emission is inhomogeneously broadened with an estimated FWHM of around $12\,$GHz (see Supplementary Material)~\cite{Kuhlmann2013,Nawrath2021}. This spectral mismatch between the bandwidth of the memory and the total inhomogeneous bandwidth of the single photon emission means that only a small proportion of the QD photons are on resonance with the memory, leading to a low memory efficiency (see Supplementary Material). In order to increase the observed memory efficiency we spectrally filter the photons from the QD single-photon source using a Fabry-Perot etalon with a bandwidth of $1.12\pm 0.02$~GHz, which selects the photons that are on resonance with the memory. The transmission of the {broadended QD emission line} through the etalon is around 3\%. The spectrum of the QD photons {filtered by} the etalon is shown in Figure~\ref{fig:QDcharacterisation}(a), and the $g^{(2)}$ measurement is shown in Figure~\ref{fig:QDcharacterisation}(b) from which we extract $g^{(2)}(0) = 0.325 \pm 0.008$. {This $g^{(2)}$-measurement shows some blinking which originates from the spectral filtering of the emission line and therefore leads to artificial on-off times of the single-photon source.}

The loss introduced by both the temporal and spectral filtering significantly reduces the average photon number per pulse at the input to the memory. The detected count rate of the single photons transmitted through the entire set-up when the control field is turned off is $(203.4 \pm 0.8) \, \mathrm{s}^{-1}$ (for a full loss budget see Supplementary Material). Whilst the telecom ORCA quantum memory has the lowest noise floor of any atom-based quantum memory~\cite{Thomas2022}, a small amount of noise is generated by the interaction of the strong control pulses with the atomic ensemble (see Supplementary Material). The etalon is placed after the memory in order to further suppress this residual noise. The etalon therefore has two purposes: to spectrally filter the quantum dot photons to match the bandwidth of the memory, and to filter any noise from the memory that is not at the signal frequency. The temporally-filtered photons from the quantum dot source are sent through the quantum memory, and the retrieved light on the output is filtered using the etalon and then sent to the detection set-up. The full experimental set-up is outlined in Figure~\ref{fig:setup}(c), and details are given in Methods.

\begin{figure}
\centering\includegraphics[width=\linewidth]{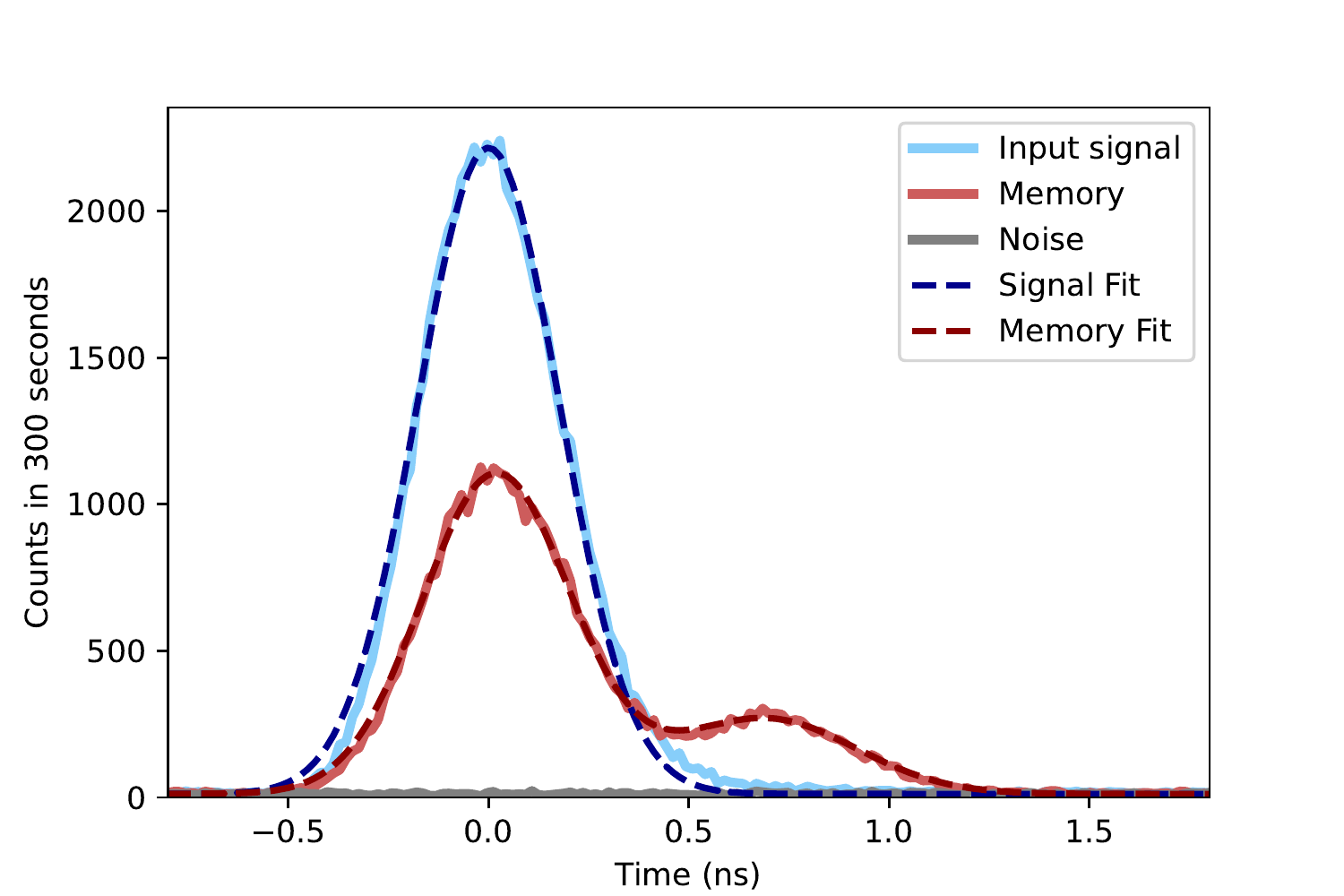}
\caption{\textbf{Storage and retrieval of QD photons in ORCA memory.} Arrival time histograms of QD photons through the ORCA memory for the input (blue), memory (red) and noise (grey) settings.  The dashed lines are Gaussian fits to the data.  \label{fig:memorydata}}
\end{figure}

Figure~\ref{fig:memorydata} shows the arrival time histogram of photons at the output of the memory. When the control fields are present (red data) we see absorption of the photons into the memory. We then see deterministic retrieval of the light in a later time window due to the application of the read-out control pulse, which is chosen here to be 800~ps later. In order to extract the efficiency of the storage and retrieval processes we fit Gaussian wavepackets to the data - one Gaussian for the input signal, and a sum of two Gaussians with a time separation for the transmitted (not stored in the memory) and retrieved signals (see Supplementary Material). From the fit to the data we extract a read-in efficiency of  $\eta_\mathrm{in} = (49.3 \pm 0.4)\% $ and a total (internal) memory efficiency of $\eta_\mathrm{tot} = (12.9 \pm 0.4)\% $.  The detected count rate of photons retrieved from the memory in the output integration window with a width of 500~ps is $(22\,\pm\,1)  \, \mathrm{s}^{-1} $. The grey data in Figure~\ref{fig:memorydata} shows the noise generated by the memory and detectors, measured with the control fields present and the input signal blocked. We note that the background count rate is the same regardless of whether the control field is turned on or not, which indicates that the noise is predominantly due to detector dark counts. We define the signal-to-noise ratio (SNR) as the ratio of signal photons retrieved from the memory compared to the number of noise photons in the output window, and we measure SNR = $18.2 \pm 0.6$. This is the first demonstration of storage and on-demand recall of photons from a solid-state single-photon source in a quantum memory.

While the retrieved count-rate is too low to measure the $g^{(2)}(0)$ of the output of the memory, it can be inferred with the measured input $g^{(2)}_\mathrm{in}(0)$ and retrieved SNR, $S$. We can model the output field as an incoherent sum of the retrieved photons and noise coming from the dark counts of the detectors. The predicted $g^{(2)}_\mathrm{out}(0)$ of the output is given by (see Methods): 

\begin{equation}
    g^{(2)}_\mathrm{out}(0) = \frac{ 1 + 2S  + S^2 g^{(2)}_\mathrm{in}(0) }{( 1 + \mathrm{S})^2} \label{eq:g2}
\end{equation}

\noindent With $g^{(2)}_\mathrm{in}(0) = 0.325\pm 0.008 $ and $S = 18.2 \pm 0.6$ we predict $g^{(2)}_\mathrm{out}(0) = 0.393 \pm 0.007 $. This indicates that the noise level of this memory is sufficiently low to allow for storage and retrieval of quantum states of light well below the non-classical threshold.  This analysis does not take into account that the origin of  $g^{(2)}_\mathrm{in}(0) \neq 0 $ could be due to insufficient suppression of the excitation laser which leads to additional unwanted photons in the input state that are distinguishable from the desired single photon contribution. These unwanted photons would therefore not be on resonance with the quantum memory and would transmit straight through, and for that reason the $g^{(2)}_\mathrm{out}(0)$ of the light that is stored and then retrieved by the quantum memory may in fact be lower than the input state. {Furthermore, the indistinguishability of photons retrieved from the memory is predicted to be higher than the input since the memory acts as a coherent mode filter~\cite{Gao2019}.} 

\section{Discussion}

This first-of-its-kind demonstration of on-demand recall of QD single photons from an atomic memory is the first crucial step towards hybrid quantum light-matter interfaces for scalable quantum networks. In order to increase the overall throughput and efficiency of this interface, further improvements are needed to better match the temporal-spectral mode overlap between the source and memory. 

To improve the spectral overlap, the inhomogeneous bandwidth of the solid-state emitters can be reduced by using electrically-contacted semiconductor devices that can significantly decrease the charge noise~\cite{Nowak2014,Zhai2020}. Alternatively, the bandwidth of the memory can be increased by increasing the control pulse bandwidth, although in order to maintain low noise levels it may be necessary to further detune from resonance requiring higher control pulse energies to maintain the same light-matter coupling efficiency. In either of these cases, the requirement of spectral filtering could be relaxed leading to significantly higher end-to-end efficiencies. 

Temporal mode filtering of QD emission was a requirement in this current demonstration due to the Doppler-limited memory lifetime of around 1~ns. This can be overcome in two ways. On the one hand, optimised QD microstructures can be used to further Purcell-enhance the emission leading to temporally shorter photons. On the other hand, the Doppler limit can be removed and the storage time of the memory increased. This can be achieved by operating the memory in a cold atomic ensemble, or by using velocity-selective pumping~\cite{Main2021}, although achieving the high optical depths required for efficient light-matter interactions with these methods represents a significant technical challenge. An alternative approach could be to introduce additional fields to map the ORCA coherence to longer-lived states with a backward read-out to facilitate Doppler rephasing \cite{Moiseev2001}, or to compensate for dephasing with controlled dynamic AC-Stark shifts \cite{Finkelstein2021}. %Alternatively, changing the atomic states used to operate the ORCA memory with infra-red wavelengths for both control and signal results in a Doppler-limited storage time in excess of $100\,$ns~\cite{Finkelstein2018}.
With a longer storage time, the full emitted single photon pulse could be mapped into the memory, thereby removing the need for lossy temporal filtering. Finally, temporal shaping of the control pulses could be used in order to optimise the storage efficiency for the input QD photon temporal mode, and in principal near-unit memory efficiency could be reached~\cite{Gao2019,Guo2019}. 

In conclusion, we have demonstrated the storage and on-demand retrieval of single photons emitted from a semiconductor quantum dot in a rubidium vapour-based quantum memory, with operation in the low-loss telecommunication bands. Over a $500\,$ps integration window we achieved a signal-to-noise ratio of $18.2 \pm 0.6$, limited only by detector dark counts, and a total memory efficiency of $(12.9 \pm 0.4)\% $. This represents a significant step towards the goal of an efficient hybrid interface between solid-state single-photon sources and atomic quantum memories, pivotal for developing future quantum technologies.

\vspace{5mm}

\section{Methods}

\small

\textbf{QD single-photon source fabrication.} 
{The QD sample structure follows the thin-film InGaAs metamorphic buffer structure introduced by Sittig et al.~\cite{Sittig2022}. The MOVPE-grown sample features 23 AlAs/GaAs DBR pairs on top of the GaAs wafer substrate followed by the jump-convex-inverse InGaAs metamorphic buffer (MMB). The non-linear variation of In content simultaneously allows for sufficient strain reduction, a smooth surface and a low MMB thickness. The MMB changes the lattice constant so the self-assembled InAs QDs emit at around 1550$\pm$75~nm. The thin MMB and the InGaAs capping layer form a $\lambda$-cavity around the QDs to benefit emission at the desired 1529.3~nm. The upper part of the sample consists of an oxide top DBR of 4 pairs SiO$_2$/TiO$_2$ with a nominal thickness of 268~nm/168~nm, and a 20~nm cap of SiO$¬_2$, to protect the surface, which was fabricated using plasma-enhanced sputter deposition. The planar cavity was designed with a strong asymmetry in the top DBR and bottom DBR reflectivities, such that the light will be guided predominantely upwards. The bottom DBR has a calculated reflectivity of 99.2\% whereas the top DBR has a calculated reflectivity of 97.2\%. {To achieve a cavity at around 1529~nm,} the thickness of the InGaAs MMB as well as the capping layer and the DBR layers have been decreased in order to shift the cavity to the lower desired wavelength. 

\textbf{QD single-photon source operation.}
The quantum dot sample is placed inside a low-vibration closed-cycle cryostat (Attocube attoDRY800) and cooled to 4~K. The sample is mounted on top of a piezo stack that allows for precision positioning along three axes. An objective lens with a focal length of 3.1~mm (Thorlabs) and a numerical aperture of NA = 0.68 is mounted inside the cryostat directly above the QD sample, which focuses the excitation laser to a spot size of approximately 2~$\mu$m on the top surface of the sample. The pulses for exciting the QD are generated from an optical parametric oscillator (OPO) which frequency converts pulses from a Titanium Sapphire (Ti:Sapph) laser at 810~nm to around 1505~nm using a non linear optical crystal. The output pulses of the OPO are around 200~fs in duration. We spectrally filter these pulses using a diffraction grating and a pinhole to narrow the spectrum to 0.9~nm or a pulse duration of 3~ps. 

The emission from the sample is separated from the excitation laser using three  12~nm FWHM bandpass filters (Thorlabs), each of which have a transmission of 95\% at 1529~nm and a suppression of approximately $10^5$ at 1505~nm. The emitted light is coupled into a single-mode fibre and sent to a spectrometer with a resolution of 5.8 GHz per pixel, and detected using a InGaAs CCD camera (Oxford Instruments Andor iDus). The wavelength of the QD emission is compared to the target wavelength of the quantum memory. 

We measure the count rate of the single-photon source on a superconducting nanowire single photon detector (SNSPD), which has an efficiency of $\eta_\mathrm{detector} = 0.80$.  To measure the single-photon purity of the collected light we use a 50:50 fibre beam splitter and send both outputs of the beam splitter to a SNSPD. We record the correlations between the two detectors using a time tagger {TT,} Swabian Instruments) to perform a second-order autocorrelation. From the correlation histogram we can extract $g^{(2)}(0)$ from the ratio of the coincidence peak at zero time delay compared to the {Poissonian-level} at long time delays (see Supplementary Material). 

We measure the temporal profile of the photons by performing a coincidence measurement between the SNSPD and the trigger from the OPO laser using the time tagger. The temporal profile of the photons is quite long compared to the maximum storage time of the quantum memory. We temporally filter the collected light using an EOM intensity modulator (iXblue), which has a transmission of 45\% and an extinction ratio of 20~dB. The EOM is driven by an RF signal from an arbitrary waveform generator (Tektronix). The trigger signal from the OPO is used as both an 80~MHz clock input and a trigger signal for the AWG, so that the intensity modulation is synchronised with the single photon generation. The AWG outputs a Gaussian RF signal with a FWHM of 300~ps. The time-filtered temporal profile of the photons is shown in Figure~\ref{fig:QDcharacterisation}(c) in the main text.

\textbf{Quantum Memory.} 
The quantum memory is based on the telecom ORCA protocol in an ensemble of warm rubidium atoms~\cite{Thomas2022}. The ORCA memory uses three atomic energy levels, $\ket{g}, \ket{e}$ and $\ket{s}$ in a ladder configuration (see Figure~\ref{fig:setup}(a)). Two counterprogating fields are both detuned from the intermediate transitions $\ket{g} \rightarrow \ket{e}$ and $\ket{e} \rightarrow \ket{s}$ by $\Delta$, but are in two-photon resonance with the $\ket{g} \rightarrow \ket{s}$ transition. The strong control field maps the incoming signal onto a coherence between the ground and storage states ($\ket{g}$ and $\ket{s}$). On application of a second control pulse at a later time, the atomic coherence is mapped back to an optical field, and the signal is retrieved from the memory. 

The implementation of the telecom ORCA memory uses an 8~cm long cell of rubidium atoms that is warmed up to around 120$^\circ$C, generating an optical depth of around 8500. The control pulses that drive the memory are generated from a mode-locked Ti:Sapph laser, which is synchronised to the OPO that excites the QD via a Lock-to-Clock module (Spectra Physics). An unbalanced Mach-Zehnder interferometer is used to generate two pulses separated by 800~ps to act as the read-in and read-out pulses of the memory. The read-in pulses have an energy of 0.4~nJ and the read-out pulses have an energy of 4~nJ. The frequency of the control pulses is monitored using a High Finesse wavemeter and is set to be 6~GHz red-detuned from the $\ket{g} \rightarrow \ket{e}$ transition. The single photons and control pulses travel counter-propagating through the vapour cell and the two fields are combined and separated using dichroic mirrors (Semrock) that transmit 1529~nm and reflect 780~nm light. The control pulses (single photons) are focused to a beam waist diameter of 250~$\mu$m (220~$\mu$m) at the centre of the vapour cell respectively. The  single photons that are transmitted or stored/retrieved from the memory are coupled into a single-mode fibre, and a series of long-pass and band-pass filters (Thorlabs/Semrock) before the fibre is used to suppress the control field by 14 orders of magnitude. More details about the experimental set-up of the memory are given in~\cite{Thomas2022}. 

The light that is collected at the output of the memory is detected using an SNSPD with 70~ps timing jitter (PhotonSpot). A time {tagger (Swabian Instruments)} is used to generate correlation histograms between the single photon detector and the trigger of the OPO laser. An arrival time histogram is measured for three settings: (i) ``Signal'': input photons only; (ii) ``Memory'': input photons and control pulses; (iii) ``Noise'': control pulses only. In order to minimise the effect of any slow drifts in the experiment, data is recorded for each of the three settings for 60 seconds in a repeated cyclic mannor, until the total cumulative counts are sufficient to extract the efficiency and signal-to-noise ratio.

\textbf{Predicted $g^{(2)}$ of output.} We model the output of the memory as an incoherent admixture of two fields: the photons that are stored and then retrieved from the memory, and the noise photons that are added. The number of photons in each field is $N_\mathrm{sig, (noise)}$ respectively, and the signal-to-noise ratio on the output is given by $S = N_\mathrm{sig}/ N_\mathrm{noise}$. The $g^{(2)}(0)$ of the resultant field is given by:

\begin{IEEEeqnarray}{rCl}
    g^{(2)}_\mathrm{out}(0) &=& \frac{N_\mathrm{sig}^2 g_\mathrm{sig}^{(2)}(0) + 2 N_\mathrm{sig} N_\mathrm{noise} + N_\mathrm{noise}^2 g_\mathrm{noise}^{(2)}(0)}{(N_\mathrm{sig}+N_\mathrm{noise})^2} \\
    &=& \frac{S^2 g_\mathrm{sig}^{(2)}(0) + 2 S +  g_\mathrm{noise}^{(2)}(0)}{(1+S)^2}
\end{IEEEeqnarray}

We set $g^{(2)}_\mathrm{sig}(0)$ to be equal to that of the input signal photons $ g^{(2)}_\mathrm{input}(0)$ since, other than this addition of the noise field we expect the light to be unchanged by the memory. The dominant noise source is from dark counts of the detectors which have multi-mode thermal statistics with  $g_\mathrm{noise}(0) = 1$. We then reach Equation~\ref{eq:g2} given in the main text. 

\bibliography{QD-QM}

\bibliographystyle{unsrtnat}

\vspace{3mm}

\textbf{Acknowledgements} -- This work was supported by Horizon 2020 research and innovation program under Grant Agreements No.899814 (QUROPE) and No.899587 (Stormytune), and by the Engineering and Physical Sciences Research Council via the Quantum Computing and Simulation Hub (Grant No T001062). The authors gratefully acknowledge the funding by the German Federal Ministry of Education and Research (BMBF) via the project QR.X (No.16KISQ013). SET acknowledges an Imperial College Research Fellowship. PML acknowledges support from UK Research and Innovation (Future Leaders Fellowship, Grant Reference MR/V023845/1). THL acknowledges support by the BMBF (Qecs - Grant No. 13N16272) and SH acknowledges support by the BMBF (QR.X - Grant No. 16KISQ010). We thank Margit Wagenbrenner for reliable layer deposition for the dielectric DBR. 

\textbf{Author Contributions} -- SET and LW performed the storage and retrieval experiments with the support of PB. SET implemented the atomic memory with help from PML, PB, SSS. LW, RJ and CN characterised the QD sample, which was grown by RS. THL and SH designed and prepared the oxide DBR. Support on the measurements, sample preparation and characterisation was provided by MJ, PM, IAW, SLP and PML. SET analysed the data and wrote the manuscript with support from all authors. PM, IAW, SLP and PML designed the experiment and coordinated the project. All authors contributed to scientific discussions. \hypertarget{equalcontributions}{$\dagger$ SET and LW contributed equally to this work.}

\clearpage
% \onecolumngrid 
	
	\appendix
	\setcounter{figure}{0} \renewcommand{\thefigure}{S.\arabic{figure}}
	\setcounter{equation}{0} 
	\renewcommand{\theequation}{S.\arabic{equation}}

\begin{center}
    \large{Supplementary Material}
\end{center}	

\section{QD Spectra}
Figure~\ref{fig:QDSpectra}(a) shows the spectrum of the planar cavity structure of the QD sample, measured using high power above-band excitation, and shows the optimisation for operation at the target wavelength of 1529.3~nm (dashed red line). Due to the large objective-NA (0.68) the cavity shows resonances for lower wavelengths at higher collection angles~\cite{Bjork1993}.  Figure~\ref{fig:QDSpectra}(b) shows an overlay of micro-PL spectra of  different quantum dots from the same sample, when excited via p-shell excitation. Several quantum dots emit at, or close to, the target wavelength of the quantum memory (1529.3~nm, dashed red line). 

\begin{figure}[h]
\centering\includegraphics[width=\linewidth]{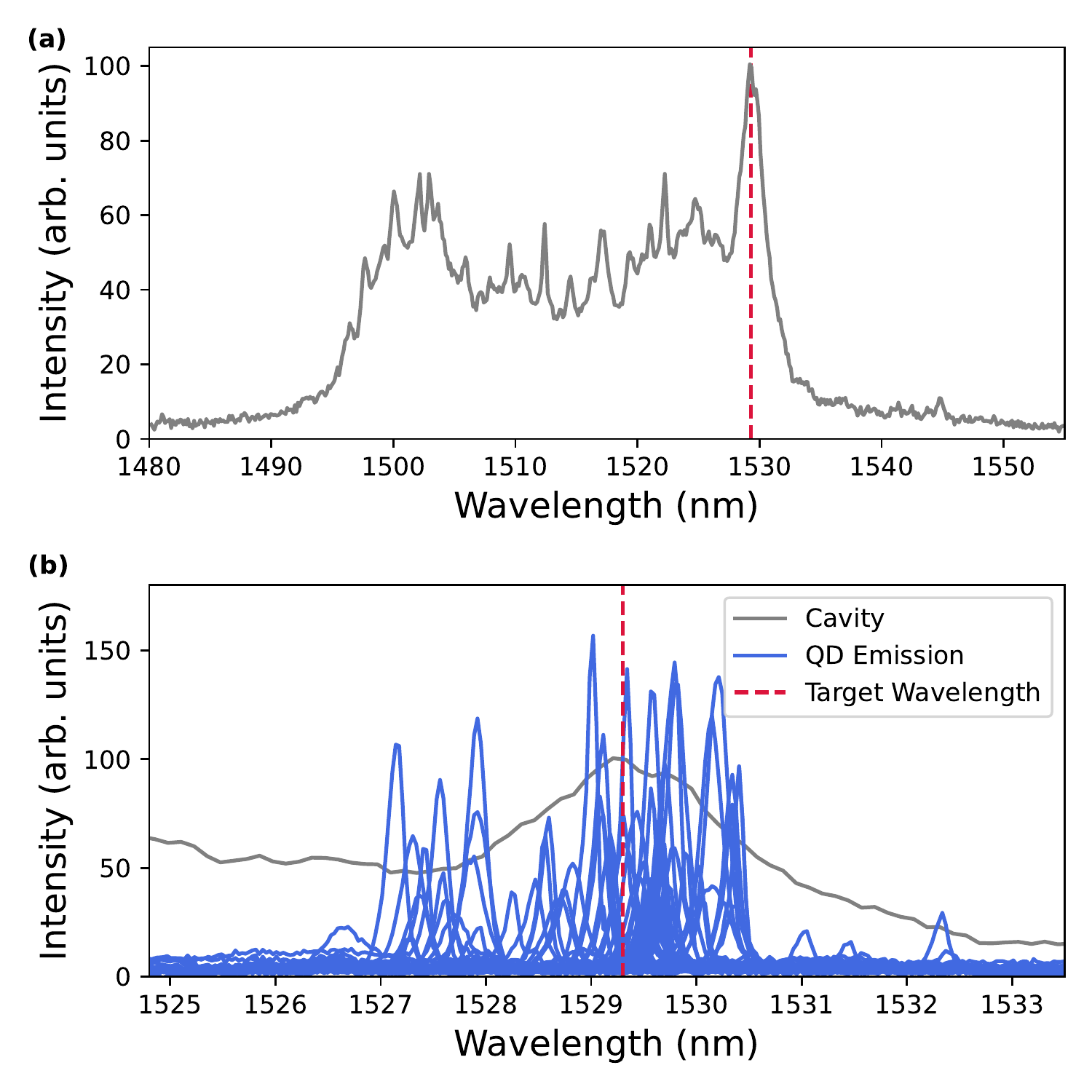}
\caption{ (a) The spectrum of the planar cavity of the QD sample, measured using high power above band pumping. The dashed red line indicates the target wavelength of 1529.3~nm (b) The {blue spectral peaks originate from individual QDs of the same sample,} measured under p-shell excitation, which emit close to the target wavelength (red). The cavity spectrum is again shown in grey, plotted over this smaller wavelength range. \label{fig:QDSpectra} } 
\end{figure}

\section{Characterising QD single-photon source}

\begin{figure}
\centering\includegraphics[width=0.9\linewidth]{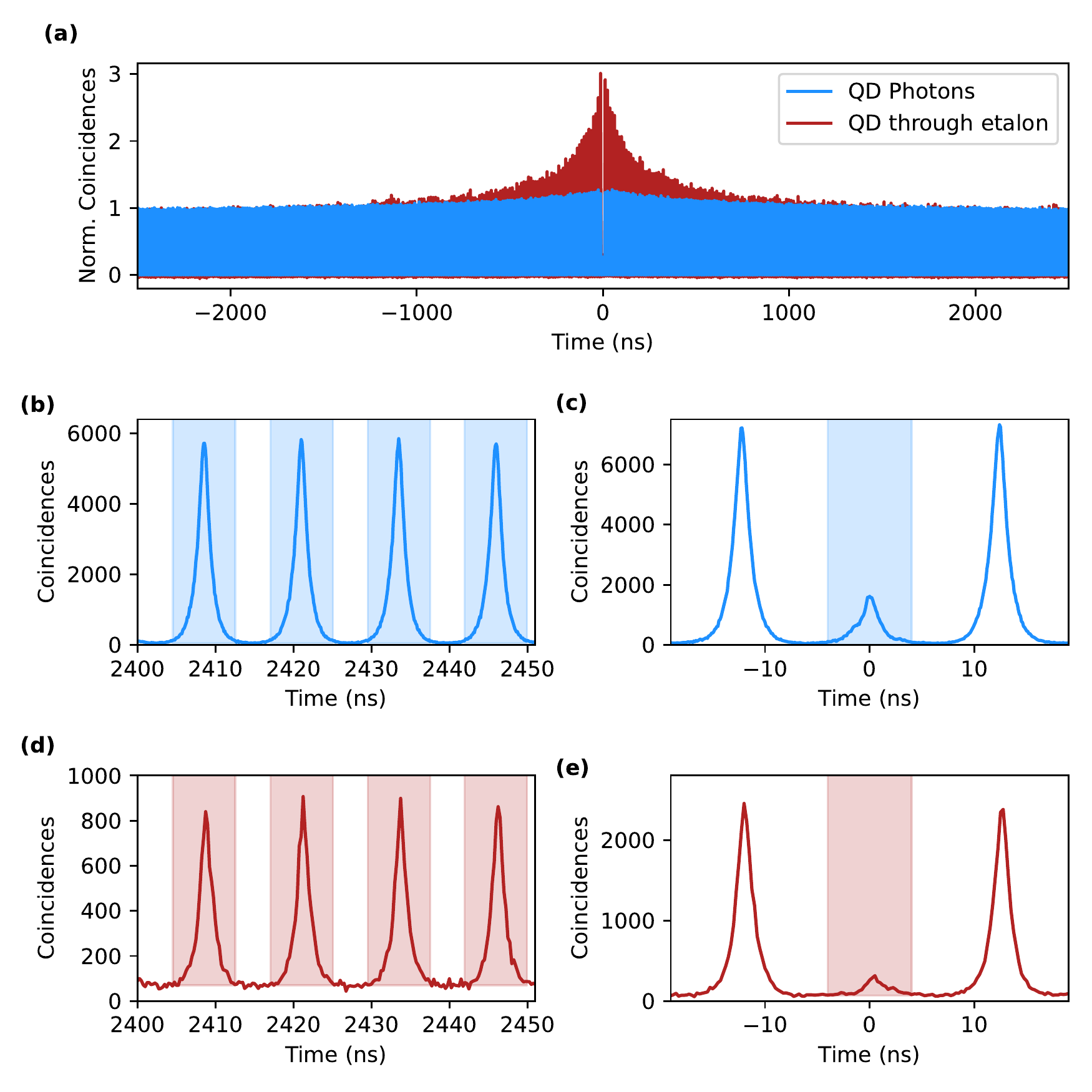}
\caption{ (a) Second-order autocorrelation of the quantum dot single-photon source without (blue) and with (red) spectral filtering from an etalon. The data is normalised to the height of Poissonain peaks at long time delays. For the data with the etalon, blinking is clearly visible. The panels (b,d) display the measurements at long time delays while panels (c,e) show the behaviour at around zero time delay. The shaded regions indicate the integration areas used to calculate the average area of the Poissonian peaks $A_\mathrm{uncor}$ (b,d), and the area of the central peak $A_0$ (c,e), in order to extract $g^{(2)}(0)$. \label{fig:g2dataanalysis}}
\end{figure}

To quantify the multi-photon contributions of the QD source we send the collected light through a 50:50 fibre beam splitter and measure the temporal correlations between the two outputs. Figure~\ref{fig:g2dataanalysis} shows the data for the correlations between the two detectors as a function of the time delay between them. The blue data is for the photons directly from the QD single-photon source, and the red data is when the collected light is spectrally filtered by a Fabry-Perot etalon. Both measurements are normalised to the Poissonian level at long time delays. When the etalon is used we see a decrease in the height of the peaks with increasing time difference between the detectors. This is because the etalon has effectively transformed the spectral wandering of the source into blinking (artificial on/off times of the source). The second-order autocorrelation at zero time delay is measured as {$g^{(2)}(0) = A_0 / A_\mathrm{Poisson}$}, where $A_0$ is the area of the peak at zero time delay, and $A_\mathrm{Poisson}$ is the average area of the {Poissonian}
peaks at long time delays. The shaded regions in Figure~\ref{fig:g2dataanalysis}(b-d) indicate the integration windows with a width of 8~ns that are used to count the area of the peaks. The extracted values are: $g^{(2)}(0)  =  0.306 \pm 0.002$ without the etalon and $g^{(2)}(0) = 0.325 \pm 0.008 $ with the etalon.

\section{Quantum Memory Noise}
\begin{figure}
\centering\includegraphics[width=0.9\linewidth]{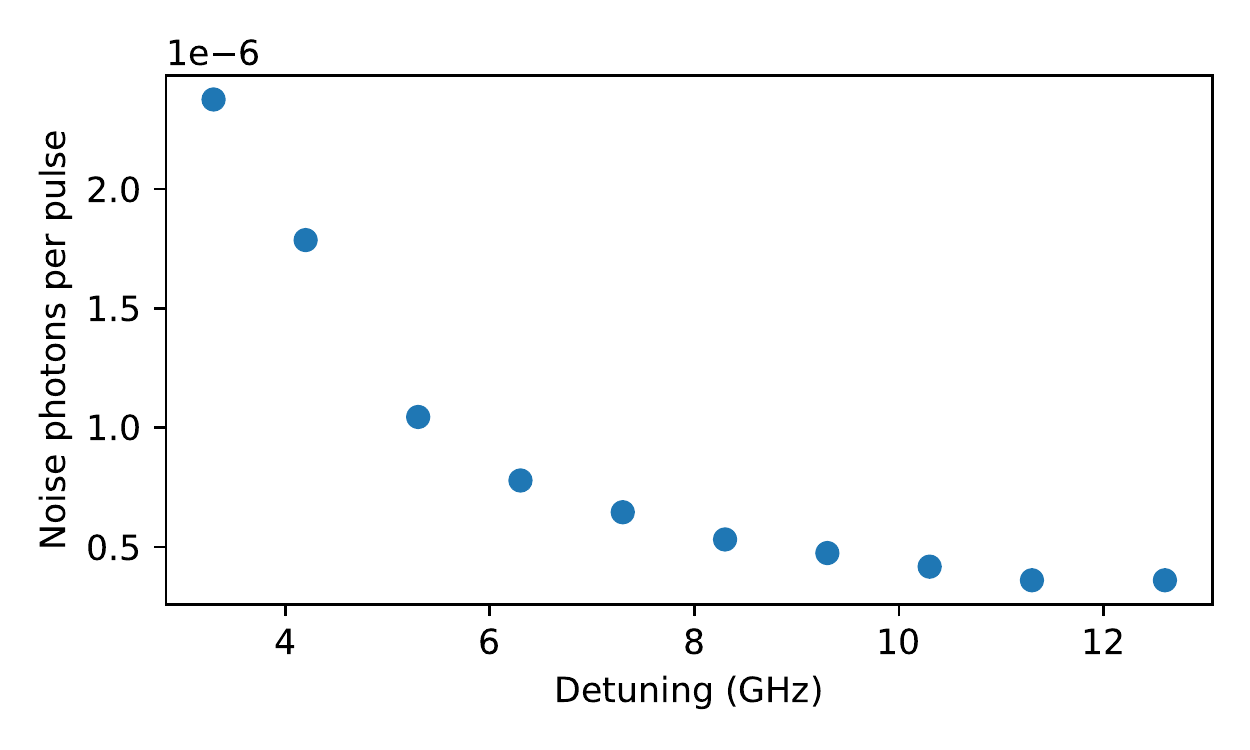}
\caption{ The noise photons per pulse generated by the ORCA memory as a function of the detuning of the control field from resonance, when no etalon is used for filtering. For all other data the control field has a detuning of 6~GHz.  \label{fig:NoiseVsDetuning}}
\end{figure}

The noise that is generated by the telecom ORCA memory with no etalon is on the order of $10^{-6}$ photons per pulse. The origin of this noise could be either direct leakage of the control field due to insufficient spectral filtering, or fluorescence noise due to the interaction of the control field with the atomic vapour. We measured the noise per pulse as a function of the detuning of the control pules from resonance, as shown in Figure~\ref{fig:NoiseVsDetuning}, and we see that the noise increases as the control field gets closer to resonance. This indicates that the noise is due to the interaction of the control field with the atoms, rather than direct leakage of the control field. However, it is not clear what the precise mechanism is that causes this fluorescence noise at telecom frequencies due to the application of the control field at 780~nm. 

When no etalon is used, the number of noise photons that we detect in a 500ps integration window is $(1.33 \pm 0.01) \times 10^{-7}$ per pulse. When the etalon is placed after the ORCA memory, the number of detected noise photons in the output integration window is  $(1.5 \pm 0.1) \times 10^{-8}$ per pulse, indicating that the spectral filtering from the etalon significantly reduces the noise.

\section{Memory performance with and without spectral filtering}

The emitted photons from the quantum dot single-photon source have a time-averaged spectrum that is larger than the memory bandwidth due to inhomogeneous broadening. We investigate storage of the single photons in the memory both with and without spectral filtering. 

We first investigate storage of the single photons into the memory without any spectral filtering, and the data is shown in Figure~\ref{fig:MemoryDataSM}(a). We measure a read-in efficiency of $\eta_\mathrm{in} = (10.16\pm 0.05)\%$, and a total (internal) memory efficiency, $\eta_\mathrm{tot}= (1.00 \pm 0.05)\%$. The read-in efficiency is low due to the spectral mismatch between the overall spectrum of the QD photons and the acceptance bandwidth of the ORCA memory which is determined by the 1~GHz control pulse. Note that for this measurement the room lights were switched on and therefore the noise is due to an elevated background count rate of the detectors. 

We then spectrally filtered the photons using an etalon with a FWHM of $(1.12 \pm 0.02)$~GHz, in order to better match the bandwidth of the photons to that of the memory. Figure~\ref{fig:MemoryDataSM}(b) shows the memory data when we spectrally filter the photons \textit{before} the memory. The read-in efficiency here is $(43.4 \pm 0.1) \%$ and the total (internal) memory efficiency is $(8.9 \pm 0.1)\%$. The signal-to-noise ratio on retrieval is $1.73 \pm 0.01$. The data in Figure~\ref{fig:MemoryDataSM}(c) is with the etalon placed \textit{after} the memory to spectrally filter both the single photons and any noise from the memory that is not at the signal frequency. The read-in efficiency here is $(49.3 \pm 0.4)\%$ and the total (internal) memory efficiency is $(12.9 \pm 0.4) \%$, indicating that the memory efficiency is similar whether the etalon is placed before or after the memory. However, the signal-to-noise ratio here is significantly higher at $18.2 \pm 0.6$.

We can use the data in Figure\ref{fig:MemoryDataSM}(b-c) to estimate the  inhomogeneous bandwidth of the un-filtered single-photon source.  We numerically solve the Maxwell-Bloch equations of the ORCA memory {for a simplified three level system, including the full velocity distribution of the atoms, using a fourth order Runge-Kutta method to increment along the time axis and a Chebyshev spectral method for solving the spatial dimension}. We first simulate the memory for the case \textit{with} spectral filtering, where we assume that both the signal and control field have a bandwidth of $1\,$GHz {and that both fields are in two photon resonance with the storage state for an atom with zero velocity}. Using a control pulse energy of $0.4\,$nJ and the measured optical depth of $8500$, we estimate a read-in efficiency of  $49.3\%$, which matches well with the measured value of $(49.3 \pm 0.4 )\%$. We then model the inhomogeneous broadening of the single-photon source by {repeatedly simulating the memory interaction while varying the central frequency of the photons input to the memory, according to a normal distribution of some width to be determined.  We find that a normal distribution with a FWHM of $12.4\,$GHz results in an average read-in efficiency of $10.12\%$, comparable with the measured value of $(10.16\pm 0.05)\%$ when no spectral filtering was applied to the photons.} Typical linewidths observed for these QDs under p-shell or resonant excitation are around 5-10~GHz~\cite{Nawrath2019}. From these numerical simulations we estimate that the chosen QD which emits at the memory frequency without tuning happens to have a slightly larger linewidth of approximately 12~GHz.

\begin{figure}
\centering\includegraphics[width=0.9\linewidth]{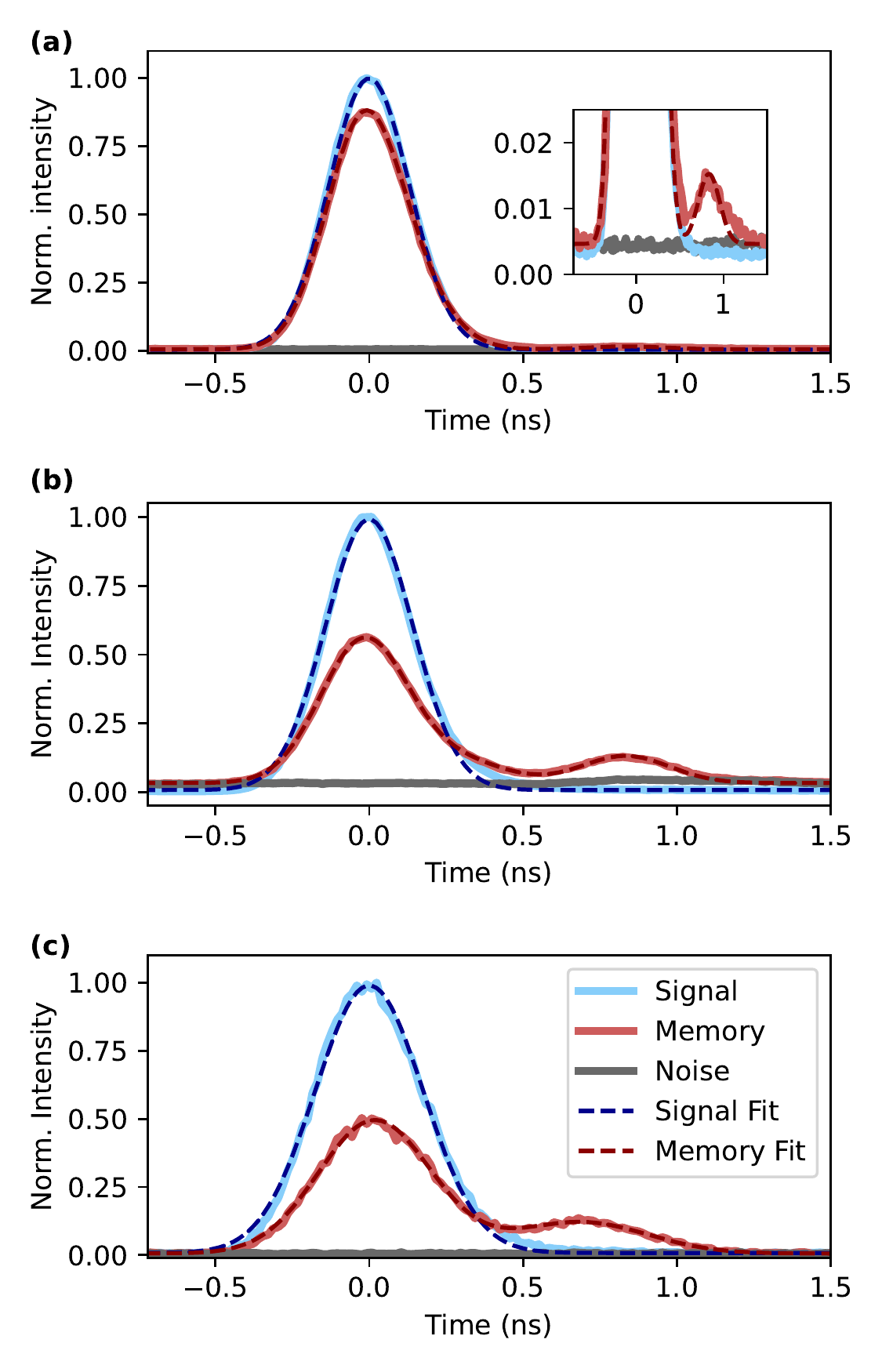}
\caption{ Arrival time histograms of QD photons through the ORCA memory for the input (blue), memory (red) and noise (grey) settings.  The three plots indicate three different experimental configurations: (a) no frequency filtering of QD photons, (b) when an etalon is used to filter the QD photons \textit{before} the memory, and (c) when an etalon is used to filter the QD photons and noise \textit{after} the memory. The dashed lines are a Gaussian fit to the data. The inset in (a) is zoomed in to show the retrieved light.  \label{fig:MemoryDataSM}}
\end{figure}

\section{Extracting Memory Efficiency and SNR}

Figure~\ref{fig:memorydatafit}(a) shows the raw data for the arrival time histograms for storage of photons from the QD in the quantum memory, with the etalon placed after the memory. The inset shows that the background counts (away from the pulses) are the same for all three settings, independent of whether the control field is on or not, indicating that the noise counts are from the dark counts of the detectors and not from the quantum memory itself. We calculate the average noise counts across the entire time trace and use this as a constant background offset for the fits to the data. 

\begin{figure}
\centering\includegraphics[width=\linewidth]{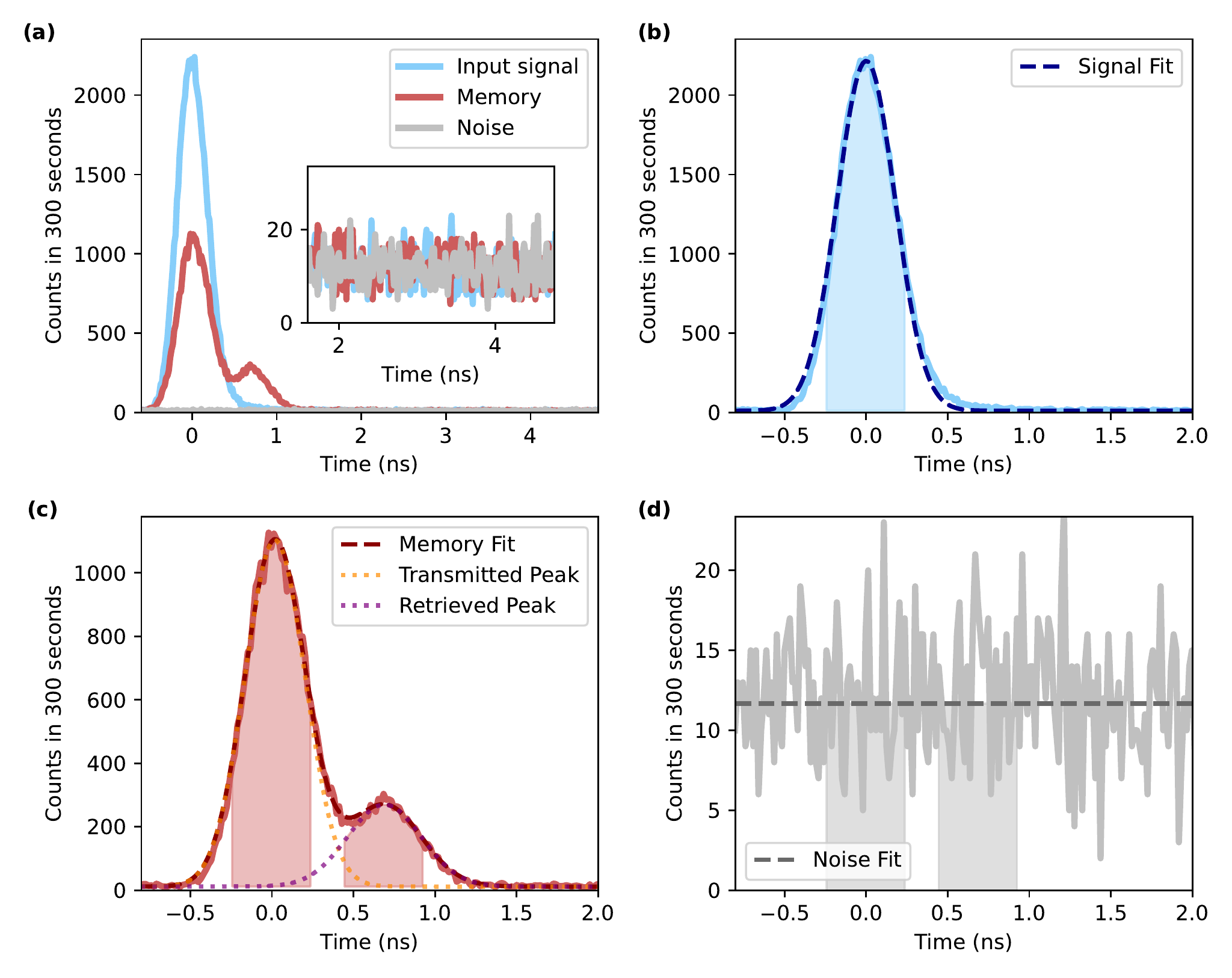}
\caption{ (a) The raw data for the storage of photons from the QD single-photon source in the quantum memory, with the etalon placed after the memory. The inset shows the number of counts away from the peak is the same for all three settings of the memory measurement. The fits to the data are shown for (b) the input signal, (c) the memory, and (d) the noise. The shaded regions indicate the areas that are taken to calculate the efficiencies and signal-to-noise ratio. The integration window here is 500~ps.  \label{fig:memorydatafit}}
\end{figure}

The data are fit with Gaussian functions of the general form
\begin{equation}
A\,\textrm{exp}\left[ - \frac{(t - t_c)^2}{2\sigma^2}\right] + \textrm{Offset}\label{eq:Gauss}
\end{equation}
with the amplitude $A$, centre $t_c$ and width $\sigma$ as free parameters with the offset fixed as stated earlier. For the case of the memory histograms, a sum of two Gaussian functions is used. This approach allows us to treat the transmitted and retrieved pulses independently, since they are measured to be slightly overlapping, thereby removing ambiguity in the calculation of the efficiencies and SNR - see Figures~\ref{fig:memorydatafit}(b-d) for the data, fits and integration windows used. The read-in efficiency is given by one minus the ratio of areas between the transmitted and input pulses, and the memory efficiency is the ratio of the retrieved and the input pulses. Finally, the SNR is the ratio of the retrieved pulse to the number of noise counts in the output integration window. 

To calculate these areas, we use the analytical form of the integral of a Gaussian:
\begin{equation}
\textrm{Area} = f(A,\sigma,t_\mathrm{int}) = 2\,\sqrt{\frac{\pi}{2}}\,A\,\sigma\,\textrm{erf}\left(\frac{t_{\mathrm{int}}/2}{\sqrt{2}\sigma}\right)\label{eq:Area}
\end{equation}
where $\textrm{erf}$ is the error function and $A$ and $\sigma$ defined as before in equation \ref{eq:Gauss}. $t_\mathrm{int}$ is the integration window width which is divided by 2 in the argument of the error function since it's defined as the integral of a Gaussian from time $t=0$ up to $t = t_\mathrm{int}/2$, and a factor of 2 is included outside the error function to obtain the total area. Note that the vertical offset is not included in these calculations, i.e. the noise is removed. When assessing the efficiencies, taking these areas directly is sufficient, whereas for the SNR and count rate it is required to rescale this area by the binning of the time-to-digital converter (in our case $16\,$ps) to form the ratio with the noise counts in the same integration window. We can then estimate the error on the Area, $\alpha_\mathrm{Area}$, by
\begin{equation}
\alpha_\mathrm{Area}^2 =  \sum_{j = \{A, \sigma, t_\mathrm{int}\}} \left( \frac{\partial f}{\partial j}  \alpha_j\right)^2
\end{equation}
where $f$ is the function for the Area given in equation \ref{eq:Area}, $\alpha_j$ are the uncertainties on the values $j = \{A,\sigma, t_\mathrm{int}\}$ as extracted from the Gaussian fit to the data, and $\frac{\partial f}{\partial j}$ are the partial derivatives of the function f. The partial derivatives are:
\begin{equation}
\frac{\partial f}{\partial A} = \frac{f}{A}, \,\,\frac{\partial f}{\partial t_\mathrm{int}} = 2\,A\,\textrm{exp}\left[ - \frac{t_\mathrm{int}^2}{2\sigma^2}\right],\,\, \frac{\partial f}{\partial \sigma} = \frac{f}{\sigma} - t_\mathrm{int}\,\frac{\partial f}{\partial t_\mathrm{int}}.
\end{equation}
Note that the error function is the integral of a Gaussian centred on time $t=0$ and so the uncertainty on the centre of the Gaussian $\alpha_{t_c}$ is instead placed on the integration window time as $\alpha_{t_\mathrm{int}}$. The $\alpha_\mathrm{Area}$ values are then propagated through to get the uncertainties on the efficiencies, SNR and estimated $g^{(2)}_\mathrm{out}(0)$.

\begin{figure}
\centering\includegraphics[width=0.9\linewidth]{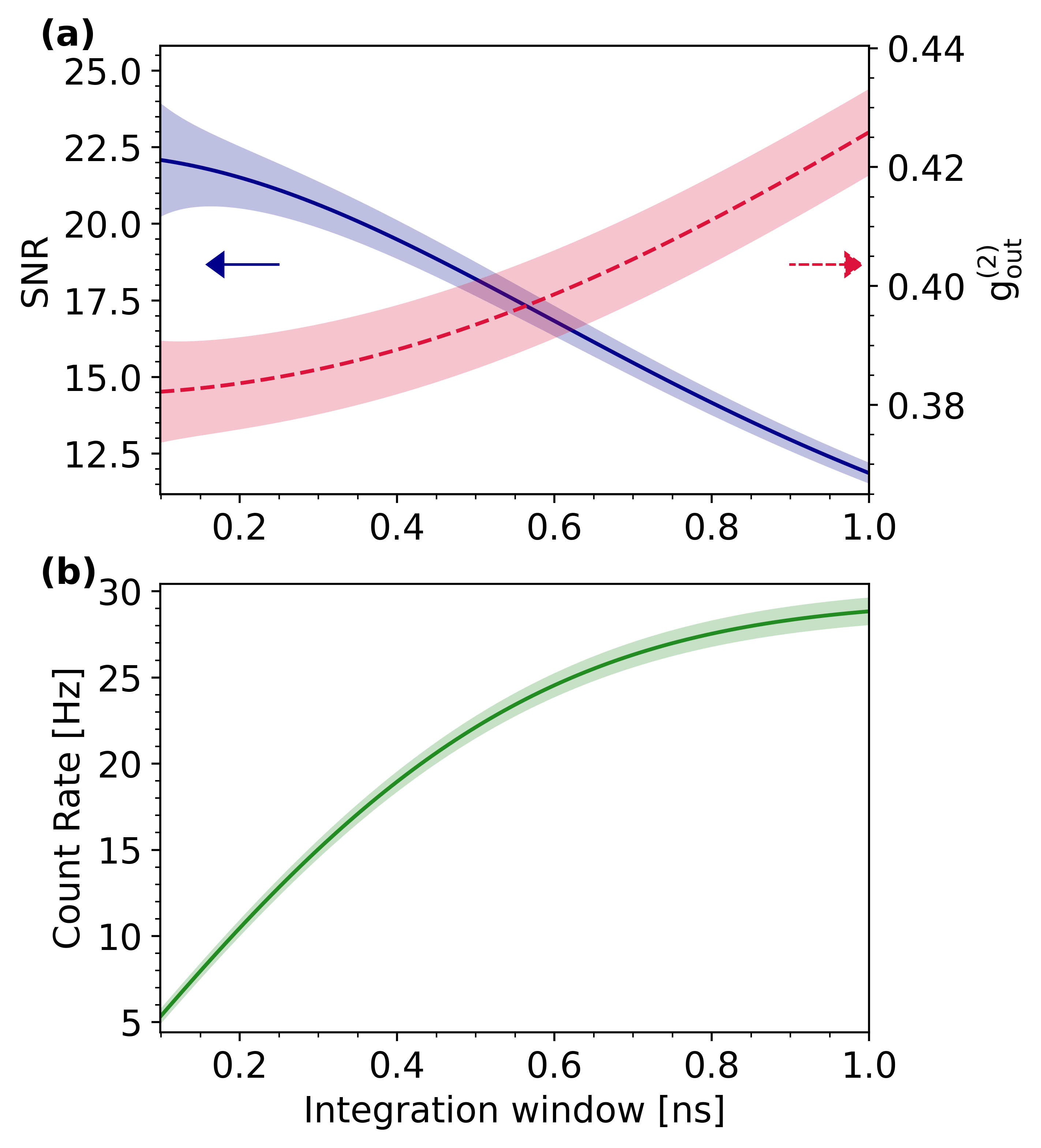}
\caption{ (a) Signal-to-noise ratio (blue, left axis) and predicted $g^{(2)}_\mathrm{out}(0)$ (red dashed, right axis) as a function of the width of the integration window.  The shaded regions indicate the uncertainty range. (b) Count rate of the retrieved light in the output integration window.  \label{fig:IntegrationWindow}}
\end{figure}

When extracting the memory efficiency and SNR we have to choose an integration window, and there is a compromise between the count rate in the output window and the signal-to-noise ratio.  Figure~\ref{fig:IntegrationWindow}(a) shows the SNR and estimated $g^{(2)}_\mathrm{out}(0)$ as a function of the integration time window, with panel (b) showing the retrieved photon count rate. We observe the typical trade-off between SNR and count rate, expected for Gaussian temporal mode signals on top of output noise that is flat over time. That is, a better SNR is observed for shorter integration windows at the expense of the overall count rate. The integration window in the main text and in Figure~\ref{fig:memorydatafit} is chosen to be 500~ps, representing a fair compromise between SNR$\,=\,18.2\,\pm\,0.6$ and count rate $\,=\,(22\,\pm\,1)\,\mathrm{s}^{-1}$. The efficiencies quoted use the same integration time window on the input pulse as is used on the transmitted and output pulses. For the chosen $500\,$ps window, about $15.5\,\%$ of the input photons are neglected. Taking the input in its entirety, with a $500\,$ps integration window for the retrieval, the memory efficiency reduces slightly to $(10.9 \pm 0.3)\%$.

\section{Loss Budget}

\begin{table}
\begin{center}
\begin{tabular}{|p{0.7\linewidth}|c|}
\hline
     \textbf{Element} & \textbf{Transmission}  \\
     \hline
     \hline
     Fibres from QD source to EOM & $ 0.90 \pm 0.01 $  \\
     \hline
     EOM Transmission & $0.45 \pm  0.02 $ \\
     \hline
     EOM output to Memory Vapour Cell (including fibre connection, dichroic mirror, waveplates, PBS, and focusing lens)  & $ 0.72 \pm 0.01 $ \\
     \hline
     Memory Vapour Cell & $0.80 \pm 0.02 $ \\
     \hline
     After Vapour Cell to Detectors (including collimating lens, bandpass and longpass filters and fibre coupling) & $0.65 \pm 0.01$ \\
     \hline
     \hline
     \textbf{Total Transmission}, $\eta_\mathrm{optical}$  & $ 0.153 \pm 0.009$ \\
     \hline
\end{tabular}
\end{center}
\caption{\textbf{Loss Budget}. The transmission of optical components in the set-up.  \label{table:lossbudget} }
\end{table}

The transmission through the set-up is characterised using a CW laser at the memory signal wavelength, $\lambda_\mathrm{S} = 1529.3$~nm, and summarised in Table~\ref{table:lossbudget}. The total optical transmission of the set-up from the single-photon source to the detectors is $\eta_\mathrm{optical} = 0.153$. The transmission from the single-photon source to the input of the memory vapour cell is 0.29. 

In addition to these transmission losses, we perform temporal and spectral filtering. The EOM is used to temporally filter the photons from the quantum dot source from a exponentially-decaying wavepacket with a $1/e$ decay time of $(0.85 \pm 0.01)$~ns to a Gaussian with a FWHM of 300~ps, which has an efficiency of $\eta_\mathrm{temporal} = 0.25 \pm 0.02$. 

A Fabry-Perot etalon is used to spectrally filter the photons. The etalon is temperature-tuned to have a peak in transmission at the signal wavelength of the quantum memory, $\lambda_\mathrm{S}=1529.3$~nm and the transmission of the CW laser at this wavelength is measured to be 85\%. The light that is transmitted through the etalon is coupled into a single-mode fibre with a coupling efficiency of 60\%. The total transmission of the CW light through the etalon and into fibre is therefore 51\%. The transmission of the inhomogenously broadened single photons through the etalon is measured to be $\eta_\mathrm{spectral} = (3.0 \pm 0.5)\%$.

\end{document}